\documentclass[12pt]{iopart}
\usepackage{graphicx}
\usepackage{amssymb}
\begin{document}
\title[The geometric calibration]{The geometric calibration of cone-beam imaging and delivery systems in radiation therapy}

\author{Evangelos Matsinos and Wolfgang Kaissl}

\address{Varian Medical Systems Imaging Laboratory GmbH, T\"{a}fernstrasse 7, CH-5405 Baden-D\"{a}ttwil, Switzerland}
\ead{evangelos.matsinos@varian.com and wolfgang.kaissl@varian.com}

\begin{abstract}
We propose a method to achieve the geometric calibration of cone-beam imaging and delivery systems in radiation therapy; our 
approach applies to devices where an X-ray source and a flat-panel detector, facing each other, move in circular orbits around 
the irradiated object. In order to extract the parameters of the geometry from the data, we use a light needle phantom which is 
easy to manufacture. A model with ten free parameters (spatial lengths and distortion angles) has been put forth to describe the 
geometry and the mechanical imperfections of the units being calibrated; a few additional parameters are introduced to account 
for residual effects (small effects which lie beyond our model). The values of the model parameters are determined from one 
complete scan of the needle phantom via a robust optimisation scheme. The application of this method to two sets of five 
counterclockwise (ccw) and five clockwise (cw) scans yielded consistent and reproducible results. A number of differences have been 
observed between the ccw and cw scans, suggesting a dissimilar behaviour of the devices calibrated in the two rotation modes. The 
description of the geometry of the devices was achieved at the sub-pixel level.\\
\end{abstract}
\noindent{\it Keywords\/}: geometric calibration\\
%\submitto{\PMB}
%\maketitle

\section{Introduction}

Featuring simultaneous tracking and targeting of tumours, image-guided radiation therapy (IGRT) paves a promising path towards 
a safer, more efficient and more accurate treatment. In autumn 2004, Varian Medical Systems, Inc.~(VMS), Palo Alto, CA, installed 
the first clinically-applicable solution for IGRT: the VMS Clinac accelerator equipped with imaging functionality.

Unless a number of calibrations have been performed on such a complex system, the device is not operable. The aim of the present 
paper is to describe one such important process, the geometric calibration. The method outlined here applies equally to imaging 
and delivery units where the source and a flat-panel detector (mounted in such a way as to face one another) move in circular 
orbits around the irradiated object. The rotation plane of the source is assumed to contain the geometric centre of the 
detector; modifications are needed in case of tilted geometry, e.g., in a C-arm configuration.

The aim of the geometric calibration is threefold. (i) To yield the values of the important parameters defining the geometry of 
the system. If an imaging unit is calibrated, these values are subsequently used in the processing of the scan data (reconstruction 
phase). If a delivery unit is calibrated, the output may lead to the suspension of the operation of the system in case, for instance, 
that the deviations of the parameter values from the nominal ones are beyond the tolerance limits. (ii) To provide an assessment of 
the deviation from the ideal world, where the machine components move smoothly and rigidly in exact circular orbits; these effects 
have appeared in the literature as `mechanical flex', `nonidealities', `nonrigid motion of the system components', etc. (iii) To 
enable the investigation of the long-term mechanical stability of the system; this examination involves the creation of a database 
containing the output of each calibration.

On top of the arguments which were just put forth concerning the necessity of the geometric calibration separately for imaging and 
delivery units, there is one additional remark which specifically applies to IGRT systems: the geometric calibration links together 
the two units comprising the machine. With the relationship between the two components of the system having been set, the updated 
anatomic information (obtained from the imaging unit) may be quickly processed and the dose distribution re-evaluated in the area 
which is subsequently subjected to radiation (delivery unit); thus, modifications in the treatment plan are enabled on the daily 
basis, reflecting the most up-to-date information in the region of interest.

Concerning earlier methods pertaining to the calibration of the geometry in computed tomography, an overview may be obtained from the 
paper of Noo \etal (2000); that work set forth an approach to extract the parameters of the geometry from the elliptical projections 
of fixed points, and inspired additional research in the field. Fahrig and Holdsworth (2000) employed a small steel ball bearing (BB) 
placed at the isocentre, and traced its projection across a series of images; corrections to the scan data were subsequently calculated 
(as a function of the gantry angle) from the movement of the image centroid on the detector. A similar approach was followed by Jaffray 
\etal (2002), whereas Mitschke and Navab (2003) developed a method featuring a CCD camera attached to the head of the X-ray source. 
Siewerdsen \etal (2005) use a phantom consisting of a helical pattern of BBs; the data analysis results in the creation of their 
`flex maps', which then lead to the assessment of the mechanical nonidealities. Finally, the interesting paper of Cho \etal (2005) 
introduces a phantom consisting of an arrangement of $24$ steel BBs in two circular patterns and achieves the description of the 
geometry of the system via a set of spatial lengths and distortion angles.

In this paper, we present one of the methods which are currently in use in the geometric calibration of the VMS devices; a shorter 
description of this image-based calibration technique has appeared in Matsinos (2005). Contrary to the bulky cylinders which are 
generally needed in other schemes in order to extract the parameters of the geometry, we use an easy-to-handle (i.e., to mount and 
dismount) and light needle phantom which, additionally, can easily be manufactured. The extraction of the values of the model parameters 
is achieved via a method which has been developed with attention to the robustness of the final outcome.

\section{Materials and methods}

\subsection{The needle phantom}

The needle phantom comprises five cylindrical $60$-mm-long metallic needles ($\varnothing$ $3$ mm) embedded in a urethane compound 
(Obomodulan 500, OBO-Werke GmbH \& Co.~KG, Stadthagen, Germany) of cylindrical form. The dimensions of the urethane housing are: 
$175$ mm (diameter) and $80$ mm (height). The weight of the needle phantom is about $900$ gr. The needles are made of a 
chromium-nickel alloy (1.4305, Edelstahlwerke S{\"u}dwestfalen GmbH, Siegen, Germany) and are coplanar, parallel and equidistant 
($30$-mm separation); the axis of the central needle coincides with the symmetry axis of the needle phantom. Marks have been incised 
into its surface to enable the proper alignment with a laser system.

The needle phantom may easily be mounted directly onto the VMS couch (Exact Couch); it is placed in such a way that the needles are 
parallel to the rotation axis.

\subsection{The detector}

The VMS PaxScan 4030CB amorphous-silicon flat-panel detector, currently used in the data acquisition, is a real-time digital 
X-ray imaging device comprising $2048 \times 1536$ square elements (pixels); the detector spans an approximate area of $40 \times 30$ 
cm$^{2}$. In order to expedite the data transfer and processing, the so-called half-resolution ($2 \times 2$-binning) mode is used in 
almost all applications; thus, the detector is assumed to consist of $1024 \times 768$ (logical) pixels (pitch: $388$ $\mu$m). 
The detector is connected to the body of the system via a set of robotic arms enabling three-dimensional (3D) movement.

\subsection{The devices}

\begin{figure}
\begin{center}
\includegraphics [width=7.5cm] {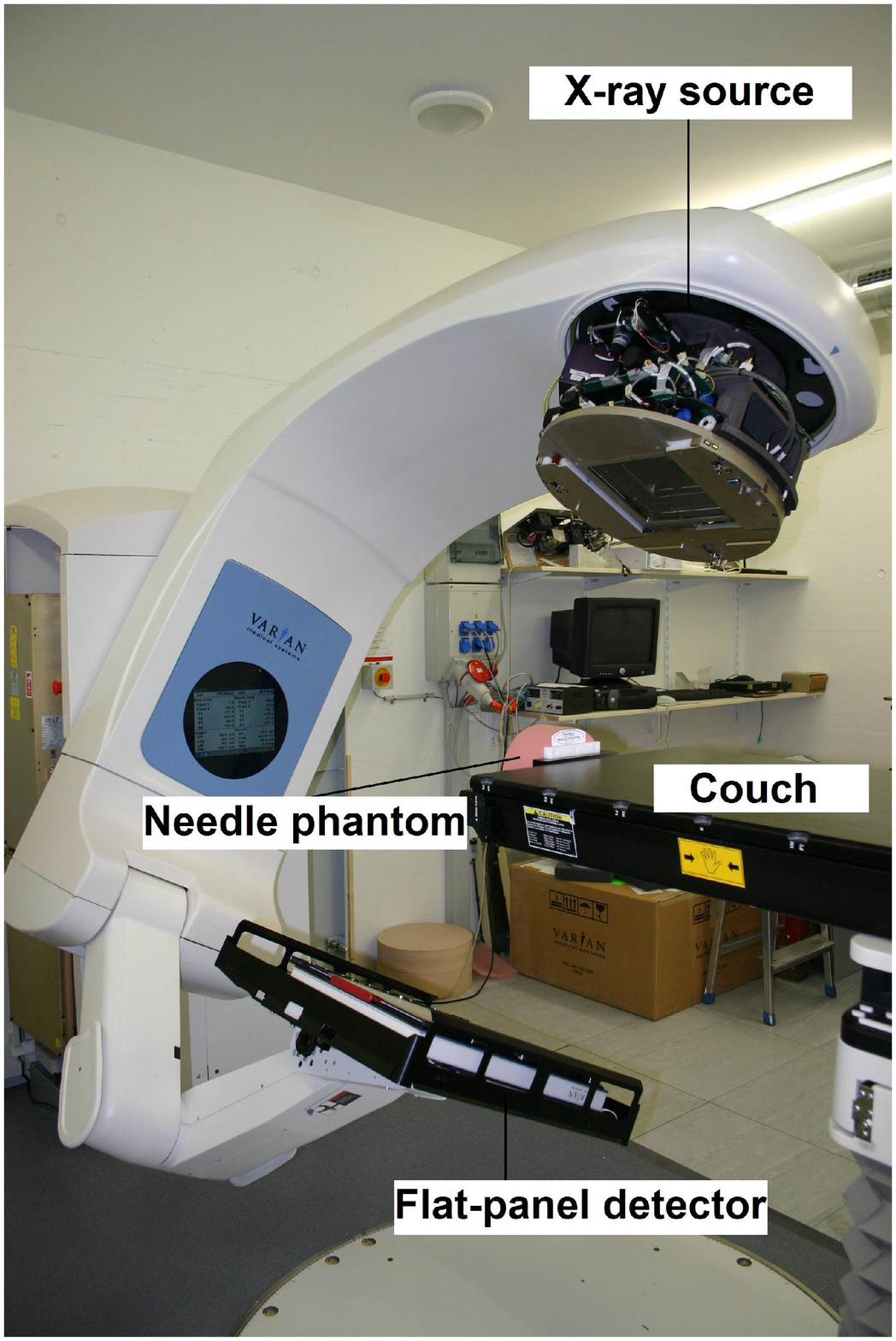}
\includegraphics [width=7.5cm] {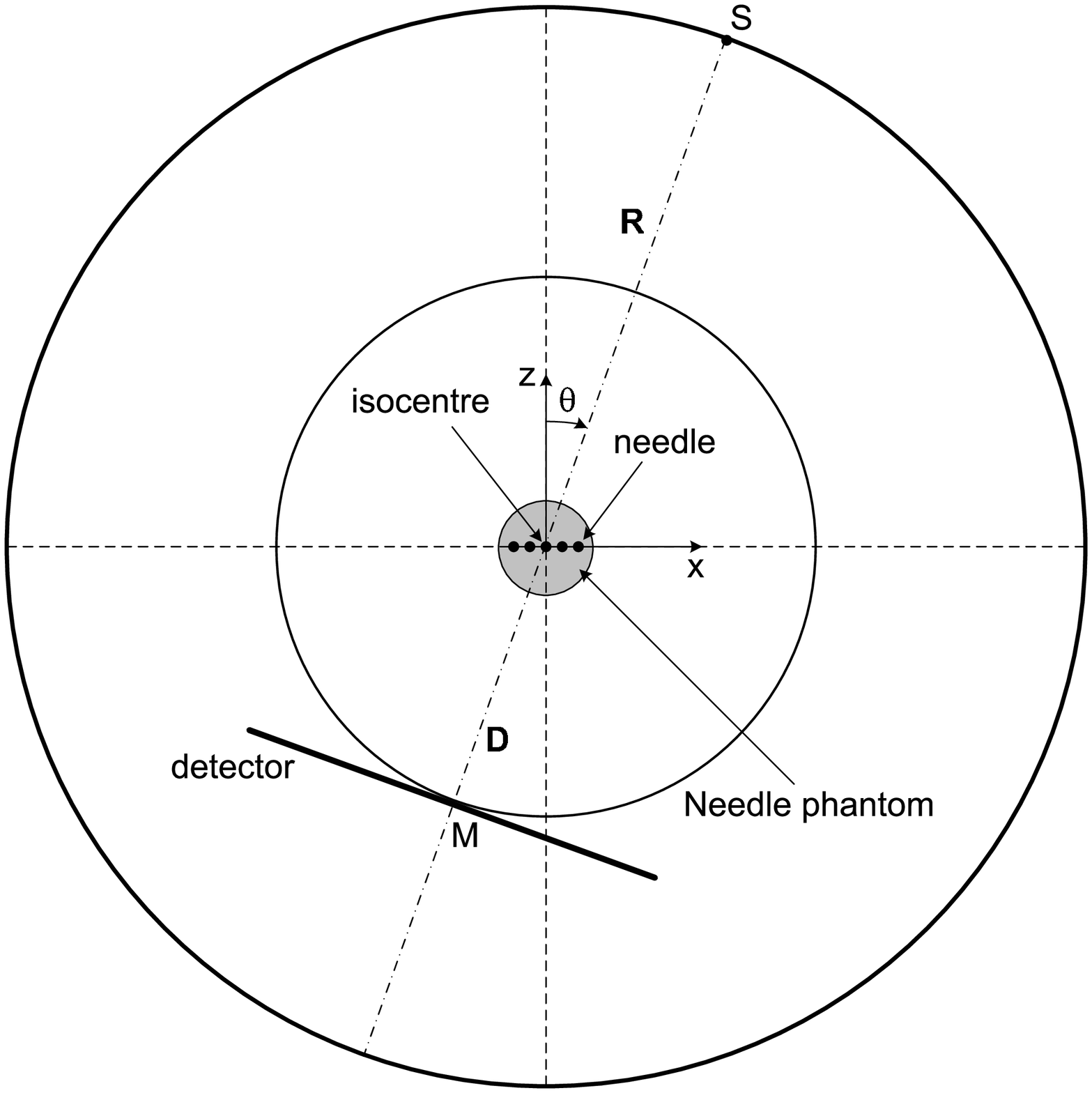}
%\vspace{-6cm}
\caption{\label{fig:PlacementAtIsocentre}Left: The VMS Acuity-Simulator device (VMS laboratory, Baden, Switzerland). The X-ray source is at a 
position corresponding to about $\theta=30^0$. Right: A schematic view of the various elements of a system on the rotation plane (facing the 
gantry, as in the figure on the left). Shown is the placement of the needle phantom in the ideal geometry; the axis of the central needle 
coincides with the rotation axis. $R$ and $D$ denote the source-isocentre and isocentre-detector distances.}
\end{center}
\end{figure}

The experimental data which are analysed in the present paper were obtained at the VMS laboratory in Baden, Switzerland, and involved 
two VMS devices: the Acuity Simulator (AS), a machine dedicated to imaging (figure \ref{fig:PlacementAtIsocentre}), and the `On-Board 
Imager' (OBI) system, the imaging unit of the VMS IGRT devices. The description of these two units may be obtained directly from the 
website of the manufacturer (`www.varian.com').

In both devices, the X-ray pulses are produced by an X-ray tube, the VMS model G242. The position of the X-ray source is fixed in the AS 
device, but adjustable (2D movement) in the OBI unit. To correct for the (large) anisotropy in the radiation field, wedge filters have 
been mounted at the head of the gantry in the AS device; there are no such filters in the OBI unit.

\subsection{The geometry}

\subsubsection{The definition of the coordinate systems.}

The ideal geometry and placement of the needle phantom are shown in figure \ref{fig:PlacementAtIsocentre}, where $R$ and $D$ denote the 
source-isocentre (often referred to as SAD) and isocentre-detector distances; $\theta$ is the gantry angle, identifying the position of 
the source. In this ideal world, the isocentre, defined as the intersection of the rotation axis and the plane on which the X-ray-source 
locus S lies during a scan, is a fixed point in space; a point source is assumed (cone-beam geometry), circumscribing an exact circle. 
The projection of the isocentre onto the detector (point M) coincides with the geometric centre of the detector; ideally, the line SM is 
perpendicular to the detector surface. Concerning the perfect orientation of the detector, two of its sides are paraller to the rotation 
plane (and two perpendicular to it). In the isocentre reference frame, a point will be represented as ($x_I$,$y_I$,$z_I$); facing the 
gantry, the $x_I$ axis points to the right, the $y_I$ axis towards the gantry and the $z_I$ axis upwards.

\begin{figure}
\begin{center}
\includegraphics [width=8cm] {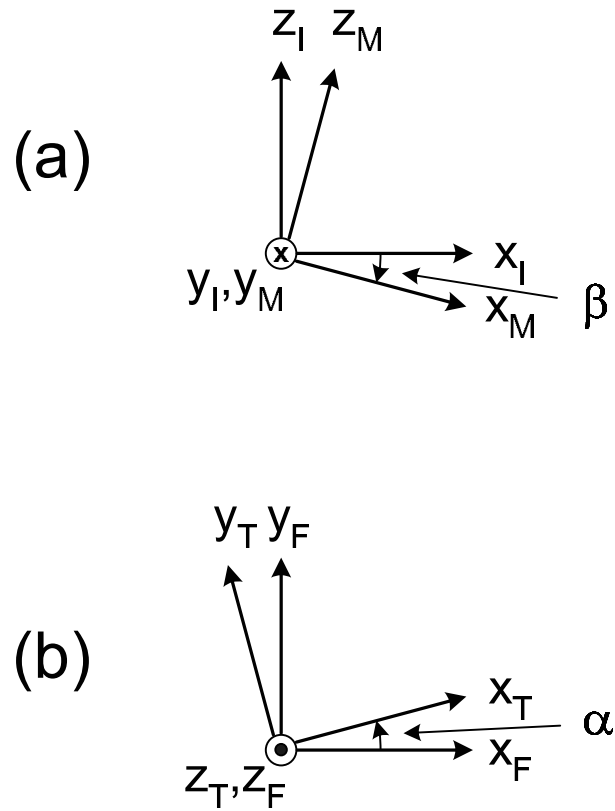}
%\vspace{-6cm}
\caption{\label{fig:CoordinateSystems}The coordinate systems used in the present paper. a) The isocentre (subscript $I$) and the 
machine (subscript $M$) reference frames; they are related via a simple rotation (angle $\beta$). b) The needle-phantom (subscript $T$) 
and the auxiliary (subscript $F$) reference frames; they are related via a simple rotation (angle $\alpha$). The auxiliary and isocentre 
reference frames are related via a simple translation involving the vector ($x_0$,$y_0$,$z_0$).}
\end{center}
\end{figure}

The second coordinate system pertains to the machine. The coordinates of points in this reference frame will carry the subscript $M$, thus, 
a point will be represented as ($x_M$,$y_M$,$z_M$). The origin of the machine reference frame is the isocentre; additionally, $y_M \equiv y_I$. 
The isocentre and machine reference frames are related via a simple rotation around the $y_I$ axis, the rotation angle being denoted as $\beta$ 
(figure \ref{fig:CoordinateSystems}a); in the ideal case ($\beta = 0^0$), the machine and isocentre reference frames are identical.

The third coordinate system is attached to the needle phantom. The coordinates of points in this reference frame will carry the subscript 
$T$, thus, a point will be represented as ($x_T$,$y_T$,$z_T$). The couch angle $\alpha$ will relate this reference frame and an auxiliary 
one (figure \ref{fig:CoordinateSystems}b), denoted as ($x_F$,$y_F$,$z_F$), having the same origin (the geometric centre of the needle phantom), 
but being parallel to the isocentre reference frame ($x_I$,$y_I$,$z_I$).

\subsubsection{The relationship of the coordinate systems.}

From figure \ref{fig:CoordinateSystems}b, it is evident that
\begin{equation} \label{eq:RelationBetweenFAndT}
\left( \begin{array}{c} x_F \\ y_F \\ z_F \end{array} \right) = \left( \begin{array}{ccc} \cos (\alpha) & -\sin (\alpha) & 0 \\ 
\sin (\alpha) & \cos (\alpha) & 0 \\ 0 & 0 & 1 \end{array} \right) \left( \begin{array}{c} x_T \\ y_T \\ z_T \end{array} \right) \, .
\end{equation}

The auxiliary and isocentre reference frames are parallel to one another. Denoting the isocentre coordinates in the auxiliary 
reference frame as $x_0$, $y_0$ and $z_0$, one obtains
\begin{equation} \label{eq:RelationBetweenIAndF}
\left( \begin{array}{c} x_I \\ y_I \\ z_I \end{array} \right) = \left( \begin{array}{c} x_F \\ y_F \\ z_F \end{array} \right) - 
\left( \begin{array}{c} x_0 \\ y_0 \\ z_0 \end{array} \right) \, .
\end{equation}

From figure \ref{fig:CoordinateSystems}a, it is deduced that
\begin{equation} \label{eq:RelationBetweenMAndI}
\left( \begin{array}{c} x_M \\ y_M \\ z_M \end{array} \right) = \left( \begin{array}{ccc} \cos (\beta) & 0 & -\sin (\beta) \\ 
0 & 1 & 0 \\ \sin (\beta) & 0 & \cos (\beta) \end{array} \right) \left( \begin{array}{c} x_I \\ y_I \\ z_I \end{array} \right) \, .
\end{equation}

These three transformations relate the sets of coordinates of a point in the needle-phantom and machine reference frames. Two parameters 
($\alpha$ and $\beta$) have been introduced to account for the miscalibration of the couch and gantry angles~\footnote{In principle, one 
additional parameter may be introduced to describe the inclination of the axes of the needles (with respect to the floor). In reality 
however, the extension of the model in this direction is superfluous. First, the couch may be considered, by all standards, parallel to the 
ground; additionally, the needle phantom is mounted onto the couch in a way which leaves no room for misplacement. Second, given the smallness 
of this departure from the ideal geometry, our results do not show any sensitivity to this degree of freedom.}. Additionally, the parameter 
$\beta$ takes account of the movement of the source between the time instant corresponding to the retrieval of the gantry-angle value from the 
system and an average time associated with the actual exposure. Including also the quantities $x_0$, $y_0$ and $z_0$, the transformation of the 
coordinates (from the needle-phantom to the machine reference frame) involves five parameters in total.

\subsubsection{The geometry in the machine reference frame.}

\begin{figure}
\begin{center}
\includegraphics [width=7.5cm] {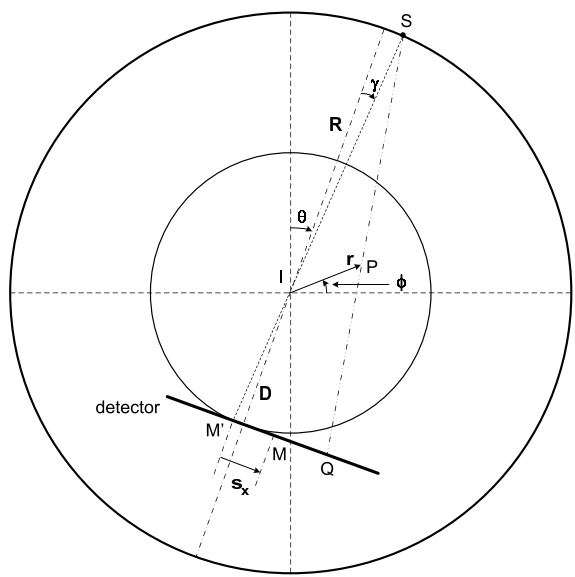}
%\vspace{-6cm}
\caption{\label{fig:DeviationsOnPlane}Machine reference frame: derivation of the projection of a point P (on the rotation plane) onto the 
detector; the quantity $r$ represents the length IP. Two deviations from the ideal geometry (on the rotation plane) have been introduced: 
the angle $\gamma$ and the lateral displacement of the detector $s_x$. The isocentre position is denoted by I.}
\end{center}
\end{figure}

The ideal geometry has already been shown in figure \ref{fig:PlacementAtIsocentre}; we will now introduce two deviations from this hypothetical 
situation \emph{on the rotation plane}. First, the straight line drawn from the isocentre perpendicular to the detector might not contain the 
source; in figure \ref{fig:DeviationsOnPlane}, this imperfection is represented by the angle $\gamma$ (positive clockwise). Second, the projection 
M$^\prime$ of the isocentre onto the detector might not coincide with the middle of the segment defined by the intersection of the detector 
and the rotation plane; the lateral (tangential to the circular orbit) displacement M$^\prime$M of the detector will be denoted as $s_x$ 
(positive to the right as one faces the gantry at $\theta=0^0$). Starting from a point P($x_M$,$z_M$) on the rotation plane and introducing 
its polar coordinates as $r$ and $\phi$, the projected length $\mid \!\! M^{\prime}Q \!\! \mid$ onto the detector is given by the formula
\begin{equation} \label{eq:MprimeQ}
\mid \!\! M'Q \!\! \mid = \left( R + \frac{D}{\cos (\gamma)} \right) \frac{r \, \cos (\theta + \gamma + \phi)}{R \, \cos (\gamma)-r \, \sin (\theta + \phi)} \, .
\end{equation}
As the geometric centre of the detector is the origin of the detector reference frame, the trace Q corresponds to a lateral reading 
equivalent to $\mid \!\! MQ \!\! \mid = \mid \!\! M'Q \!\! \mid - \mid \!\! M'M \!\! \mid = \mid \!\! M'Q \!\! \mid - s_x$.

The projected length $\mid \!\! QP' \!\! \mid$ onto the detector in the longitudinal direction (parallel to the rotation axis) for an 
abritrary point ($x_M$,$y_M$,$z_M$) is given by the formula
\begin{equation} \label{eq:QPprime}
\mid \!\! QP' \!\! \mid = \frac{R \, \cos (\gamma) + D}{R \, \cos (\gamma)-r \, \sin (\theta + \phi)} \, y_M \, .
\end{equation}
In analogy to the lateral direction, one must introduce a parameter representing the longitudinal displacement of the detector; this 
parameter will be denoted as $s_y$. Thus, the trace P$^\prime$ corresponds to a longitudinal reading equivalent to the length 
$\mid \!\! QP' \!\! \mid - s_y$.

\subsubsection{The orientation of the detector.}

The deviation in the orientation of the detector from the ideal geometry is a source of systematic effects. This misorientation may easily 
be described in terms of three rotations around the principal axes of the detector, corresponding to the lateral and longitudinal directions, 
as well as to the one which is perpendicular to its surface. Due to the fact that some of our parameters show sensitivity to the last rotation 
(i.e., around the detector normal), we will introduce one parameter ($\eta$) to account for this degree of freedom; concerning the two former 
rotations (i.e., around the lateral and longitudinal directions), our results show no sensitivity, at least up to the level at which these 
distortions are present in the VMS devices which were calibrated. On the contrary, the inclusion of the parameter $\eta$ leads to an improved 
description of the longitudinal residuals; we will address this point later on.

Denoting the coordinates of a trace on the detector as $x_d$ and $y_d$ in the nonrotated reference frame, the coordinates $x'_d$ and $y'_d$ in 
the rotated reference frame are obtained via the tranformation
\begin{equation} \label{eq:DetectorRotation}
\left( \begin{array}{c} x'_d \\ y'_d \end{array} \right) = \left( \begin{array}{cc} \cos (\eta) & \sin (\eta) \\ 
-\sin (\eta) & \cos (\eta) \end{array} \right) \left( \begin{array}{c} x_d \\ y_d \end{array} \right) \, .
\end{equation}

\subsection{The method}

Equations (\ref{eq:MprimeQ})-(\ref{eq:DetectorRotation}) establish a relationship between the coordinates $x_M$, $y_M$ and $z_M$ of an arbitrary 
point and its corresponding trace on the detector; the trace is obtained from (\ref{eq:MprimeQ}) and (\ref{eq:QPprime}) after involving the detector 
displacement (that is, the parameters $s_x$ and $s_y$) and the transformation (\ref{eq:DetectorRotation}) on the detector plane. By using also 
the transformations (\ref{eq:RelationBetweenFAndT})-(\ref{eq:RelationBetweenMAndI}), one may determine the projection of an arbitrary point onto the 
detector plane starting from its coordinates $x_T$, $y_T$ and $z_T$ in the needle-phantom reference frame. A number of parameters have been introduced 
in three steps: five parameters ($\alpha$, $\beta$, $x_0$, $y_0$ and $z_0$) are associated with the transformation of the coordinates from the 
needle-phantom to the machine reference frame; another five parameters ($R$, $D$, $s_x$, $s_y$ and $\gamma$) pertain to the geometry in the machine 
reference frame; finally, the parameter $\eta$ describes the orientation of the detector. All these parameters, save for $R$ and $D$, vanish in an 
ideal world, devoid of mechanical imperfection and inaccurate calibrations.

The input data to the geometric calibration comprise one complete scan of the needle phantom. The traces corresponding to the end-points of the 
needles are identified in the acquired images; these two coordinates (for each needle, in each input image) represent the `experimental' values. 
For a given set of parameter values, the projections of the needle end-points onto the detector plane are calculated using the chain of equations 
(\ref{eq:RelationBetweenFAndT})-(\ref{eq:DetectorRotation}); these calculated values comprise the `predictions'. An optimisation scheme is set, 
by varying the model parameters and seeking the best description of the input data; a standard $\chi^2$ function is minimised. We will now touch 
upon four issues which we consider important in our work.

\subsubsection{The detection of the needle end-points.}

\begin{figure}
\begin{center}
\includegraphics [height=10cm,angle=270] {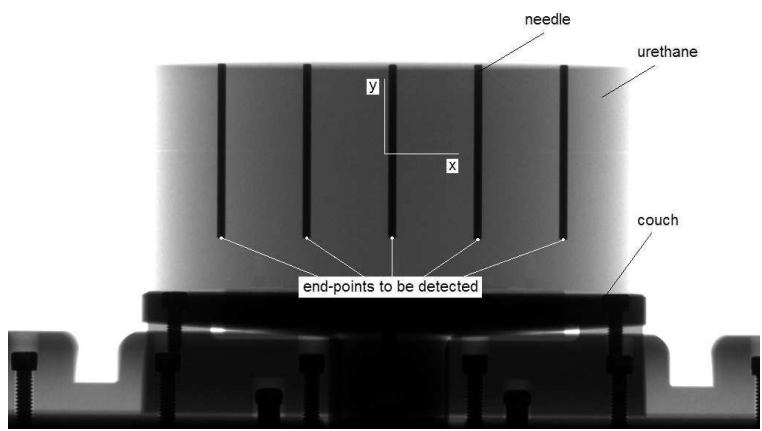}
%\vspace{-6cm}
\caption{\label{fig:NeedlePhantomProjection}An example of an image of the needle phantom. The origin of the coordinate system shown coincides 
with the geometric centre of the detector; ideally, $y$ is parallel to $y_M$ and $x$ to $x_M$ at $\theta=0^0$. The shift of the axis of the central 
needle to the right is due to the nonzero values of the parameters $x_0$ and $s_x$. A small tilt, hardly visible, is due to the nonzero values of 
the couch angle $\alpha$ and of the detector orientation angle $\eta$.}
\end{center}
\end{figure}

An image of the needle phantom is shown in figure \ref{fig:NeedlePhantomProjection}. The detection of the traces of the needle end-points is done 
in two steps.

To suppress noise, the average signal along the detector midline (along the $x$ axis) is created by averaging the contents of $16$ pixels 
in the $y$ direction, $8$ on either side of the midline. To remove the urethane background reliably, a fixed-width running window is applied to 
this average-signal data, creating the difference of the integrated signal to the (linear) background (defined by the pixel values at the window 
limits). When the window covers places where only the urethane housing is projected, the values of the transformed signal are nearly $0$. As the 
window approaches the projected axis of a needle, the transformed signal first attains a positive (local) maximum, then a negative (local) minimum 
(at the position corresponding to maximum attenuation); this signature is easily identified via simple software cuts. The algorithm is very 
efficient in detecting the signals corresponding to the needles, save for gantry-angle values around $\pm 90^0$ where the projected signals 
(of the needles) overlap.

After the signal modes (the peaks in the transformed-signal spectrum) are assigned to the needles, the projection of each needle is followed 
towards the couch (along the negative $y$ direction in figure \ref{fig:NeedlePhantomProjection}). Two signal levels are identified: one 
corresponding to the projection of the needle, one to the urethane background. The projection of the needle end-point is assumed to correspond to 
the position where the signal is equal to the geometric mean of these two values.

Two coordinates ($x'_d$ and $y'_d$) are thus determined for each needle in each input image. If more or fewer than five signals are 
detected in the lateral direction, the entire image is rejected. If a needle axis was found, but the identification of the end-point 
failed, the information relating to the particular needle is removed from the contribution of the current image to the database.

\subsubsection{Concerning the robustness of the output.}

To decrease the correlations among the model parameters in the fits, it was decided to fix the distance $R$ to the nominal value corresponding 
to the unit being calibrated. For one thing, the description of the data with variable $R$ hardly improves; for another, if $R$ and $D$ are both 
treated as free parameters, the largeness of their correlation might result in cases where the fit `drifts'.

Due to the different sensitivity of the two directions ($x$ and $y$ in figure \ref{fig:NeedlePhantomProjection}) to the model parameters, the fit 
in the lateral direction ($x$ coordinates) is performed first; this fit achieves the extraction of the values of the parameters $\beta$, $x_0$ 
and $z_0$ (associated with the relationship among the various coordinate systems), and $D$, $s_x$ and $\gamma$ (associated with the geometry in 
the machine reference frame). The remaining parameters are determined from the fit to the $y$ coordinates, which is performed with the values of 
the aforementioned six parameters fixed from the $x$-direction fit. For small $\eta$, the movement of a trace in the $y$ direction is always small, 
whereas in the $x$ direction it may be small or large, depending on the position of the point being projected; this is one of the reasons for the 
larger uncertainties in the determination of $y_0$ and $s_y$ (compared to $x_0$ and $s_x$), the other one being their strong correlation.

The MINUIT package of the CERN library, see James (1994), has been invariably used in the present paper. All output uncertainties contain the Birge 
factor $\sqrt{\chi^2/\mathrm{N}\mathrm{D}\mathrm{F}}$, which adjusts the output uncertainties for the goodness of each fit; NDF denotes the number 
of degrees of freedom. The output shows no sensitivity to the values of the model parameters which are used in the first iteration of the optimisation 
scheme.

The dependence of the parameters of our model on the gantry angle can only indirectly be assessed; in our scheme, their values are 
assumed constant within one scan. If present at a significant level, a departure from constancy will manifest itself in the creation of 
large residual effects; the point will be discussed later on.

\subsubsection{The outliers.}

Experience has shown that the presence of noise in the input data might lead to the extraction of erroneous signal modes in the transformed-signal 
spectrum; this failure rate never exceeded the $0.7 \%$ level. However, to safeguard against such cases, it was decided to precede the main 
optimisation by one performed on the data of each needle \emph{separately}; this approach is more efficient in detecting and excluding the outliers. 
Since the results obtained with logarithmic forms of the objective function are considerably less sensitive to the presence of outliers (compared 
with those of the standard $\chi^2$ optimisation), the minimisation of a simple logarithmic form has been implemented in this part of the analysis. 
The outliers are identified via a software cut corresponding to a $5\sigma$ effect for the normal distribution; other values (e.g., $3$ and $4\sigma$ 
cuts) have also been used, resulting in tiny differences in the numbers quoted here.

\subsubsection{The residuals.}

The residuals are defined as the differences between the experimental values and the corresponding predictions obtained when using the optimal 
values of the model parameters. There are three reasons why the residuals are not identically zero: (i) the statistical fluctuation (random 
noise) which is always present in measurements, (ii) the inclusion of erroneous data in the input database and (iii) the use of an insufficient 
(incomplete) model in the description of the measurements.

Concerning potential sources of systematic effects in the present work, one may recall some assumptions on which our model is based; for 
instance, it may be that (within one scan) the rotation axis is not constant, the movement of the X-ray source is irregular, the connecting 
arms are distorted under the weight of the gantry and/or of the detector, etc. Additional effects may have been introduced by approximations 
assumed in the geometry of the systems; for example, it could be that additional degrees of freedom are needed in describing the orientation 
of the detector. In any case, given the smallness of these effects (which will presently be shown), the question one has to answer is whether 
the complete description of the observations in terms of a model is called for, or it is adequate to make use of a simple model, grasp the main 
features of the geometry and attempt the empirical description of the residual effects; a number of reasons compel us to adopt the latter strategy.

\begin{figure}
\begin{center}
\includegraphics [height=7.5cm,angle=270] {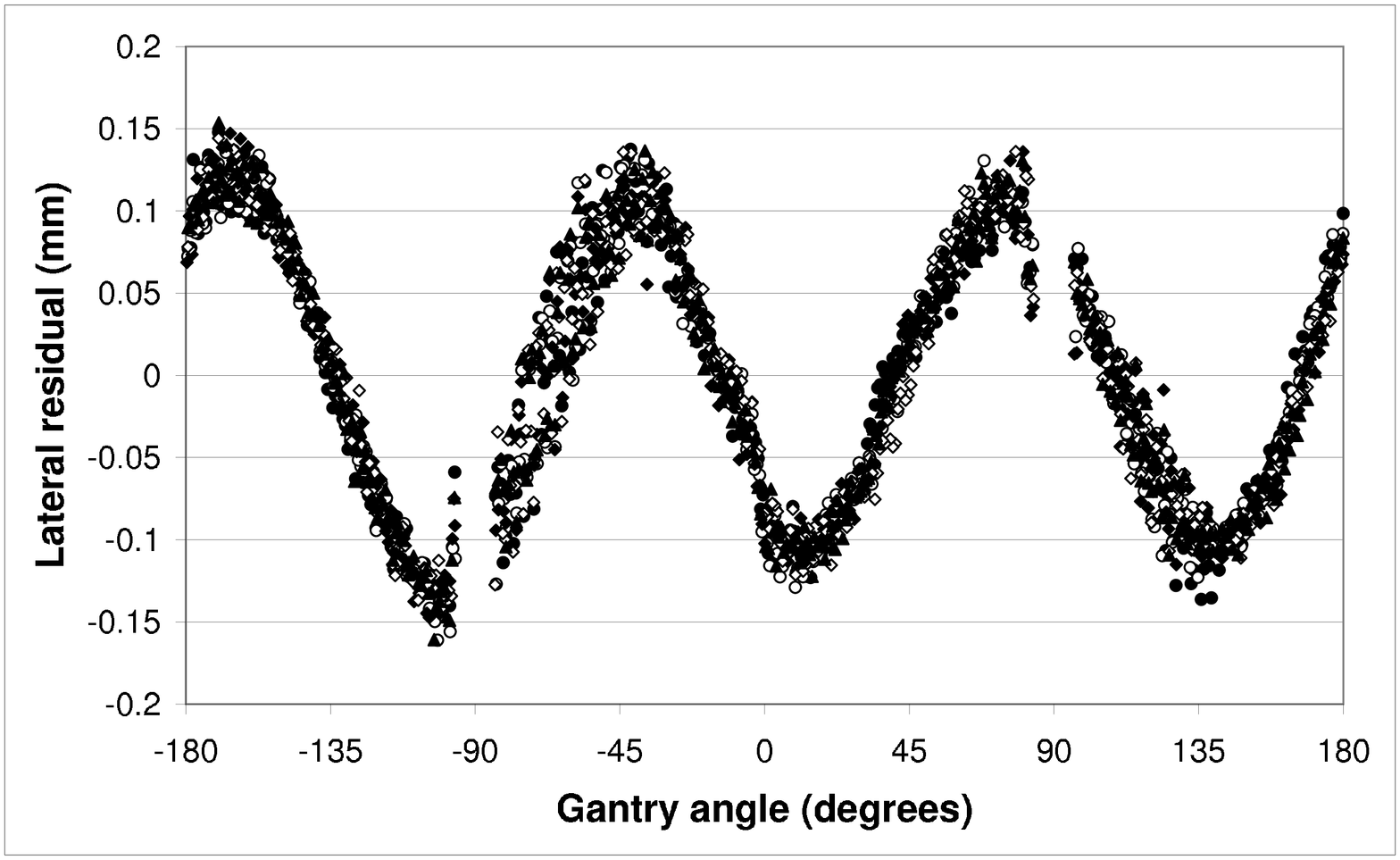}
\includegraphics [height=7.5cm,angle=270] {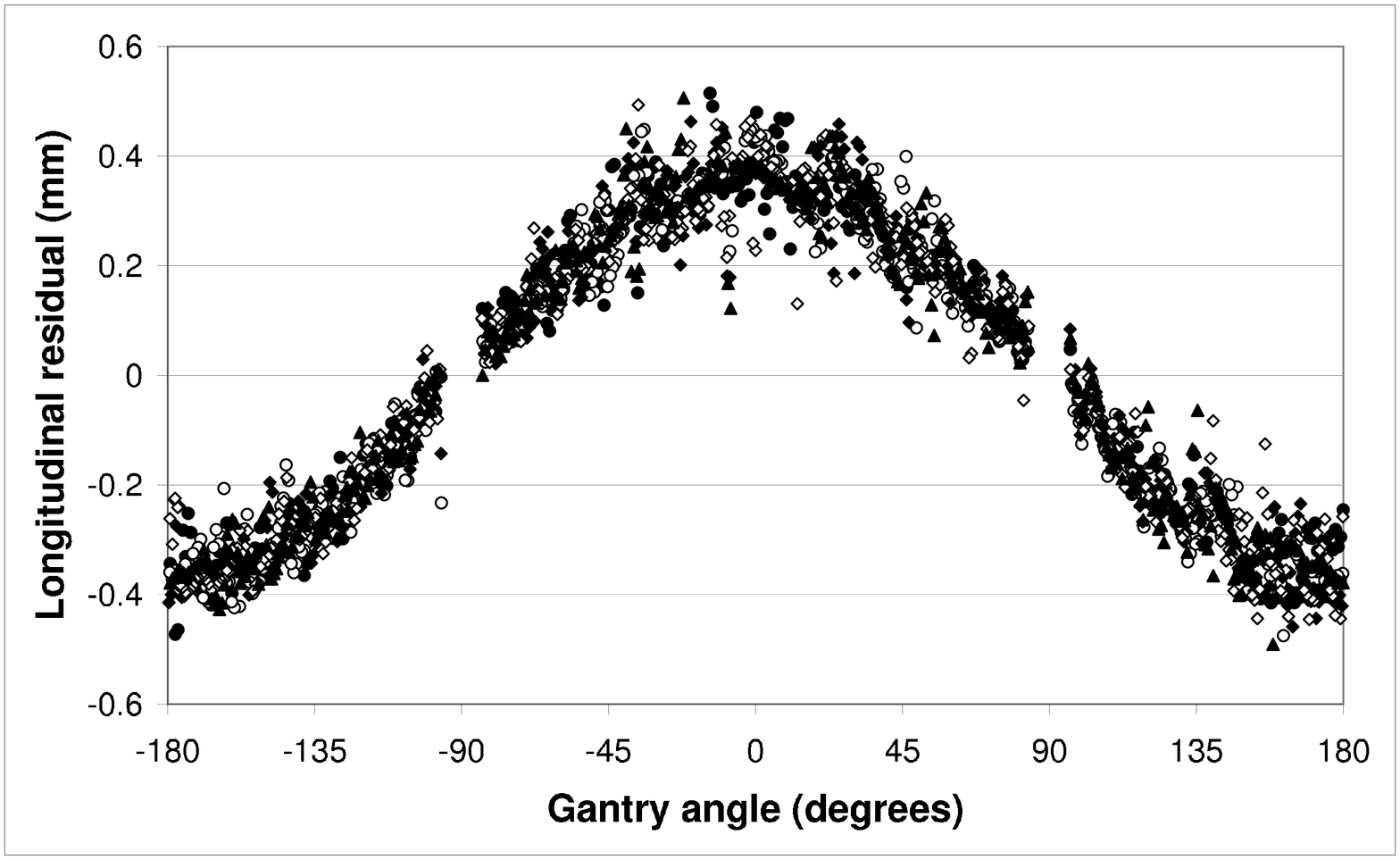}
%\vspace{-6cm}
\caption{\label{fig:Residuals}The residuals in the lateral direction (left) and those in the longitudinal direction (right) versus the gantry angle 
$\theta$ for the VMS `On-Board Imager' (OBI) unit; each point represents the average value over the needles whose end-points were successfully 
identified in the corresponding image. Shown are the results of five counterclockwise scans taken in identical conditions; these scans are 
represented by different symbols.}
\end{center}
\end{figure}

Examples of the residuals in the lateral and longitudinal directions are shown in figures \ref{fig:Residuals}. In both directions, when plotted 
versus the gantry angle separately for each needle, the residual distributions overlap; therefore, two average numbers (i.e., one per direction) 
may be used in each image. To demonstrate the reproducibility of the effect, the results of five scans taken in identical conditions are shown. 
Two conclusions can be drawn from figures \ref{fig:Residuals}: (i) the residual effects are small, the variation being of the order of one pixel 
size of the detector in the lateral and two in the longitudinal direction and (ii) the residual effects are systematic, hence they can be modelled. 
The description of the lateral ($d_x$) and longitudinal ($d_y$) residuals will be attempted by using the empirical formulae:
\begin{equation} \label{eq:ResidualsX}
d_x = A_x \, \cos (3 \, \theta + B_x)
\end{equation}
and
\begin{equation} \label{eq:ResidualsY}
d_y = A_y \, \cos (\theta + B_y) \, .
\end{equation}

\section{Results}

The experimental data were acquired on January 31, 2006. The validity of the conclusions drawn from those data was confirmed by an analysis 
of additional measurements obtained on April 26, 2006; the results of the analysis of the April data will not be given here.

The (adjustable) distance $D$ was set to $500$ mm both in the AS device and the OBI unit. The nominal $R$ value in the AS is $1000$ mm; $R$ 
was set to $1000$ mm in the OBI. On each machine, the following steps were taken. The voltage of the X-ray tube was set ($90$ kV in the AS, 
$80$ kV in the OBI). The dark- and flood-field calibrations were performed, so that flat images may be obtained in open-field geometry. The 
needle phantom was placed close to the ideal geometry of figure \ref{fig:PlacementAtIsocentre}. The acquisition settings (X-ray-tube current 
and pulse width) were chosen in such a way as to yield a good-quality signal on the detector. The frame rate was set to $8$ images per second 
on both machines, thus resulting in about $350$ images on the AS device and $450$ on the OBI unit (the rotation of the Clinac is slower). To 
investigate the short-term reproducibility of the results of the geometric calibration~\footnote{The long-term reproducibility of the resuls 
of the geometric calibration will not be addressed here. Experience has shown that, unless a device is taken apart, the geometric calibration 
does not have to be repeated more often than on a bimonthly basis.}, five counterclockwise (ccw) and five clockwise (cw) scans were acquired 
in identical geometry (except for the rotation of the gantry, no other movement of the system components was allowed) on each of the two 
machines. Both series started with a ccw scan; successive scans were taken in opposite directions.

It is noteworthy that, save for a trivial difference in the values of parameter $\beta$, no other effect (between ccw and cw scans) is expected 
in systems behaving identically in the two rotation modes. We have already mentioned that $\beta$ takes account of two effects: (i) the miscalibration 
of the gantry angle and (ii) the movement of the source during the time interval from the retrieval of the gantry-angle value (from the system) 
to the instant  associated with the `average' of the actual exposure; as the gantry-angle value is obtained a few micro-seconds after the 
`beam-on' condition (i.e., at the beginning of the emission of the radiation pulse), the time delay of case (ii) above may be thought of as 
being equal to half the acquisition setting for the pulse width. Since $6$-msec pulses were used, the difference in the $\beta$ values between 
ccw and cw scans is expected to be about $0.05^0$ in the AS device and $0.04^0$ in the OBI unit. The average of the two $\beta$ values between 
ccw and cw scans provides an estimate of the miscalibration of the gantry angle.

Our results for the optimal values of the model parameters are shown in table \ref{tab:1-table}, separately for ccw and cw scans. Inspection of 
this table leads to the following conclusions.
\begin{itemize}
\item The values of the model parameters come out reasonable. The values of the isocentre-detector distance $D$ are almost identical in ccw 
and cw scans and close to the expectation value. The values of the distortion angles are small.
\item The variation in the values of the model parameters within each rotation mode is, in most cases, smaller than the corresponding statistical 
uncertainty.
\item The $\beta$ and $\gamma$ values come out different in the two rotation modes in the AS device. We will prove, however, that these differences 
are the product of the strong correlation between these two parameters. Concerning the OBI data, the difference in the $\beta$ values may entirely 
be attributed to the delay time.
\item The largest difference between the two rotation modes in the AS device relates to the description of the residuals in the longitudinal 
direction. The two amplitudes $A_y$ of equation (\ref{eq:ResidualsY}) come out reasonably close, but the phase shifts $B_y$ do not. This 
discrepancy has been systematic for a long period of time and has been verified using other techniques; it is \emph{not} a result of strong 
correlations among the model parameters. Due to the lack of obvious structure in the distribution of the residuals in the lateral direction 
in the AS device (some structure was observed in part only of each scan), it was decided not to attempt a fit using equation (\ref{eq:ResidualsX}) 
there. The difference in the $B_y$ values for the OBI unit is statistically significant, but not so pronounced as in the case of the AS device.
\item The remaining differences are small in absolute value; for instance, the effect in $x_0$, the largest one observed, corresponds to 
one-ninth of the pixel size of the detector in the AS device and between one-sixth and one-seventh in the OBI unit. Notwithstanding the smallness 
of the mismatch, it is evident that the systems do not behave identically in the two rotation modes, suggesting the introduction of two sets of 
corrections to the scan data.
\end{itemize}

The issue of the correlation among the model parameters has to be properly addressed. The parameters $\beta$ and $\gamma$ are strongly 
correlated in the fit to the $x$ coordinates (of the traces of the needle end-points), whereas $y_0$ and $s_y$ are correlated in the 
$y$-direction fits; the remaining elements of the Hessian matrix are (in absolute value) smaller than $0.05$, indicating insignificant 
correlations. Due to the fact that the differences in the values of the parameters $y_0$ and $s_y$ in the two rotation modes are not 
statistically significant and, additionally, these two parameters are not correlated with any other, we will investigate the correlation 
only between the parameters $\beta$ and $\gamma$. From the theoretical point of view, these two parameters are independent; the distortion 
angle $\gamma$ defines the offset of the source with respect to the detector in the machine reference frame, whereas $\beta$, pertaining 
to the relationship between the isocentre and machine reference frames, applies equally to all machine components. In reality however, 
possibly coupled with the smallness of these distortion angles, a strong correlation between $\beta$ and $\gamma$ has been observed. To 
investigate whether the differences between the two rotation modes may be attributed to the correlation between these two 
parameters, we analysed the measurements acquired in the AS device further, after fixing $\gamma$ at $0^0$; the results for the model 
parameters are given in table \ref{tab:2-table}. We found that the description of the data with fixed $\gamma$ is as good as when it 
is allowed to vary freely. The changes induced by fixing $\gamma$ are entirely absorbed by parameter $\beta$, the values of which turn 
out now to be in good agreement with the expectation, based on the movement of the source within the delay time (`beam-on' condition to 
average exposure). The values of all other parameters are almost intact. Our conclusion is that the correlation between the model parameters 
$\beta$ and $\gamma$ affects only the values of these two parameters.

A good measure of the mechanical stability of a system may be obtained from the unexplained variation in the data, that is, from whichever 
fluctuation survives after all model contributions have been deducted. As previously mentioned, in the case of the AS runs, the unexplained 
variation in the lateral direction is represented by the entire fluctuation contained in these residuals, whereas in the longitudinal direction 
the empirical formula (\ref{eq:ResidualsY}) is assumed to be part of the model. For the ccw scans, the rms of the unexplained variation in 
the lateral direction is equal to $62$ $\mu$m; for the cw scans, it is $80$ $\mu$m. Therefore, the ccw scans seem to be somewhat smoother in 
the AS device. The two values of the rms of the unexplained variation in the longitudinal direction come out almost identical: $67$ $\mu$m. 
As the residuals in both directions are structured in the case of the OBI unit~\footnote{Concerning the content of the present paper, the 
most important difference between the AS and the OBI units pertains to the application of the online corrections for gravity effects; in 
the AS device, only the detector position is corrected for, whereas the corrections are applied both to the detector and the X-ray source 
on the OBI system.}, they have been fitted to via equations (\ref{eq:ResidualsX}) and (\ref{eq:ResidualsY}). Figures \ref{fig:Residuals} 
correspond to the five ccw scans in the OBI unit. In case of the ccw scans, the rms of the unexplained variation in the lateral direction 
is equal to $48$ $\mu$m; in the cw scans, it is $75$ $\mu$m. The corresponding numbers, assuming no modelling of the lateral residuals, 
are $177$ and $181$ $\mu$m; therefore, the unexplained variation in the lateral direction drops significantly when involving the empirical 
modelling of these residuals. In both rotation modes, the rms values of the unexplained variation in the longitudinal direction come out 
identical: $108$ $\mu$m.

Comparing the results obtained in the AS device with those extracted from the OBI unit, one notices that the AS scans are less noisy in the 
longitudinal direction. The values of the unexplained variation in the data in the longitudinal direction come out the same for ccw and cw scans, 
smaller than one-fifth of the pixel size of the detector in the AS device, one-third in the OBI unit. In the lateral direction, the unexplained 
variation is smaller than one-sixth of the pixel size of the detector in both systems. It is interesting to note that, after equation 
(\ref{eq:ResidualsX}) has been invoked in the description of the residuals in the lateral direction, the ccw scans on OBI correspond to a 
significantly smaller value of the unexplained variation; a similar effect is observed in the AS device, where equation (\ref{eq:ResidualsX}) was 
not used. In any case, the description of the data has been achieved at the sub-pixel level in both systems which were calibrated.

\section{Conclusions}

The aim of the geometric calibration of cone-beam imaging and delivery systems is threefold: to yield the values of important parameters 
in relation to the geometry of the system, to provide an assessment of the deviation from the ideal world (where the machine components 
move smoothly and rigidly in exact circular orbits as the gantry rotates) and to enable the investigation of the long-term mechanical 
stability of the system. The method described here applies to devices where an X-ray source and a flat-panel detector (facing each other) 
move in circular orbits around the irradiated object. Contrary to the bulky cylinders, which are generally needed in other works in order 
to extract the parameters of the geometry, we introduce a light needle phantom which is easy to manufacture.

A model has been set up to describe the geometry and the mechanical imperfections of the system being calibrated. The model contains 
five parameters associated with the transformation of the coordinates from the needle-phantom to the machine reference frame; another 
five parameters account for the geometry in the machine reference frame; finally, one parameter is introduced to account for the deviation 
in the orientation of the detector from the ideal geometry. To avoid strong correlations among the important parameters of the model, the 
source-isocentre distance is set to the nominal value of the device being calibrated. The input data comprise one complete scan of the 
needle phantom. The end-points of the needles are identified in the acquired images and comprise the `experimental' values. A robust 
optimisation scheme has been put forth to enable the extraction of the model parameters from the entirety of the input data.

The application of the approach to two sets of five counterclockwise (ccw) and five clockwise (cw) scans, acquired in two imaging devices 
manufactured by Varian Medical Systems, Inc., yielded consistent and reproducible results. The values of the model parameters come out 
reasonable. The description of the data has been achieved at the sub-pixel level.

A number of differences have been seen between ccw and cw scans, suggesting that the devices do not behave identically in the two 
rotation modes; we are not aware of other papers which have investigated and reported this effect. As a indispensable part of our 
project, we have introduced and implemented a calibration scheme in which different parameter sets apply to the scan data depending 
on the rotation mode used. We would like to draw attention to this effect, since the differences, albeit small, are systematic.

\begin{ack}
The authors would like to thank H Riem for designing the needle phantom and for obtaining the data analysed in the present paper.
\end{ack}

\References
\item[] Cho Y, Moseley D J, Siewerdsen J H and Jaffray D A 2005 Accurate technique for complete geometric calibration of cone-beam 
computed tomography systems {\it Med. Phys.} {\bf 32} 968-83
\item[] Fahrig R and Holdsworth D W 2000 Three-dimensional computed tomographic reconstruction using a C-arm mounted XRII: Image-based 
correction of gantry motion nonidealities {\it Med. Phys.} {\bf 27} 30-8
\item[] Jaffray D A, Siewerdsen J H, Wong J W and Martinez A A 2002 Flat-panel cone-beam computed tomography for image-guided radiation 
therapy {\it Int. J. Rad. Onc. Biol. Phys.} {\bf 53} 1337-49
\item[] James F 1994 {\it MINUIT, Function minimization and error analysis} (CERN: CERN Program Library Long Writeup D506)
\item[] Matsinos E 2005 Current status of the CBCT project at Varian Medical Systems {\it Proc. SPIE} {\bf 5745} 340-51
\item[] Mitschke M and Navab N 2003 Recovering the X-ray projection geometry for three-dimensional tomographic reconstruction 
with additional sensors: Attached camera versus external navigation system {\it Med. Im. Anal.} {\bf 7} 65-78
\item[] Noo F, Clackdoyle R, Mennessier C, White T A and Roney T J 2000 Analytic method based on identification of ellipse parameters 
for scanner calibration in cone-beam tomography {\it Phys. Med. Biol.} {\bf 45} 3489-508
\item[] Siewerdsen J H, Moseley D J, Burch S, Bisland S K, Bogaards A, Wilson B C and Jaffray D A 2005 Volume CT with a flat-panel 
detector on a mobile, isocentric C-arm: Pre-clinical investigation in guidance of minimally invasive surgery {\it Med. Phys.} {\bf 32} 
241-54
\endrefs

\newpage
\vspace{0.5cm}
\begin{table}%[h!]
{\bf \caption{\label{tab:1-table}}}The optimal values of the model parameters in the VMS Acuity-Simulator (AS) and `On-Board Imager' (OBI) units. 
The values quoted represent averages of five counterclockwise (ccw) and five clockwise (cw) scans. All lengths are in mm, all angles in degrees. 
The first uncertainties are systematic, the second statistical.
\vspace{0.2cm}
\begin{center}
\begin{tabular}{|c|c|c|c|c|}
\hline
	   & AS (ccw) & AS (cw) & OBI (ccw) & OBI (cw) \\
\hline
$\alpha$ & $-0.0582 (4) (13)$  & $-0.0511 (2) (13)$  & $0.2517 (2) (40)$   & $0.2550 (4) (40)$   \\
$\beta$  & $-0.169 (27) (35)$  & $0.179 (23) (38)$   & $0.112 (28) (62)$   & $0.141 (16) (64)$   \\
$x_0$	   & $1.3234 (13) (9)$   & $1.2795 (14) (11)$  & $-0.1889 (38) (17)$ & $-0.2490 (14) (17)$ \\
$y_0$    & $0.898 (11) (33)$   & $0.894 (15) (34)$   & $-0.45 (1) (10)$    & $-0.45 (0) (10)$    \\
$z_0$    & $-0.4419 (19) (10)$ & $-0.4659 (15) (12)$ & $0.2955 (60) (18)$  & $0.2806 (64) (19)$  \\
\hline
$\eta$   & $0.3101 (9) (18)$   & $0.2936 (10) (18)$  & $-0.2626 (12) (55)$ & $-0.2687 (9) (54)$  \\
\hline
$D$      & $496.670 (11) (33)$ & $496.625 (5) (38)$  & $498.401 (64) (59)$ & $498.417 (43) (61)$ \\
$s_x$    & $-2.5804 (18) (14)$ & $-2.5968 (13) (15)$ & $-1.309 (15) (2)$   & $-1.309 (16) (2)$   \\
$s_y$    & $-2.988 (13) (50)$  & $-2.978 (19) (51)$  & $-0.49 (1) (15)$    & $-0.49 (1) (15)$    \\
$\gamma$ & $0.064 (28) (35)$   & $-0.230 (22) (38)$  & $-0.103 (29) (62)$  & $-0.126 (16) (64)$  \\
\hline
$A_y$    & $0.2932 (17) (20)$  & $0.2663 (14) (20)$  & $0.3662 (25) (61)$  & $0.3568 (14) (60)$  \\
$B_y$    & $-20.26 (46) (42)$  & $-3.85 (23) (47)$   & $-0.2 (0.2) (1.0)$  & $4.9 (0.3) (1.0)$   \\
$A_x$    &                     &                     & $-0.1084 (10) (11)$ & $-0.1058 (28) (14)$ \\
$B_x$    &                     &                     & $-51.0 (1.8) (0.6)$ & $-60.8 (1.3) (0.8)$ \\
\hline
\end{tabular}
\end{center}
\end{table}

\vspace{0.5cm}
\begin{table}%[h!]
{\bf \caption{\label{tab:2-table}}}The optimal values of the model parameters in the VMS Acuity-Simulator (AS) unit for $\gamma=0^0$. The 
values quoted represent averages of five counterclockwise (ccw) and five clockwise (cw) scans. All lengths are in mm, all angles in degrees. 
The first uncertainties are systematic, the second statistical.
\vspace{0.2cm}
\begin{center}
\begin{tabular}{|c|c|c|}
\hline
	   & ccw & cw \\
\hline
$\alpha$ & $-0.0582 (4) (13)$  & $-0.0511 (2) (13)$  \\
$\beta$  & $-0.1043 (6) (14)$  & $-0.0511 (11) (16)$ \\
$x_0$	   & $1.3234 (13) (9)$   & $1.2795 (14) (11)$  \\
$y_0$    & $0.898 (11) (34)$   & $0.893 (14) (34)$   \\
$z_0$    & $-0.4419 (20) (10)$ & $-0.4659 (15) (12)$ \\
\hline
$\eta$   & $0.3081 (3) (18)$   & $0.3007 (3) (18)$   \\
\hline
$D$      & $496.669 (13) (33)$ & $496.642 (3) (38)$  \\
$s_x$    & $-2.5787 (22) (10)$ & $-2.6027 (14) (12)$ \\
$s_y$    & $-2.988 (13) (51)$  & $-2.977 (17) (51)$  \\
\hline
$A_y$    & $0.2932 (17) (20)$  & $0.2663 (14) (20)$  \\
$B_y$    & $-20.26 (47) (42)$  & $-3.85 (22) (47)$   \\
\hline
\end{tabular}
\end{center}
\end{table}

\clearpage
% ============= FIGURE1

\clearpage
% ============= FIGURE2

\clearpage
% ============= FIGURE3

%\clearpage
% ============= FIGURE5

\clearpage
% ============= FIGURE6

\end{document}